# Preventing Extinction and Outbreaks in Chaotic Populations

Frank M. Hilker[1,*] and Frank H. Westerhoff[2,†]

1. Instituto Gulbenkian de Ciência, Apartado 14, 2781-901 Oeiras, Portugal;
2. Department of Economics, University of Osnabrück, 49069 Osnabrück, Germany

abstract: Interactions in ecological communities are inherently nonlinear and can lead to complex population dynamics including irregular fluctuations induced by chaos. Chaotic population dynamics can exhibit violent oscillations with extremely small or large population abundances that might cause extinction and recurrent outbreaks, respectively. We present a simple method that can guide management efforts to prevent crashes, peaks, or any other undesirable state. At the same time, the irregularity of the dynamics can be preserved when chaos is desirable for the population. The control scheme is easy to implement because it relies on time series information only. The method is illustrated by two examples: control of crashes in the Ricker map and control of outbreaks in a stage-structured model of the flour beetle *Tribolium*. It turns out to be effective even with few available data and in the presence of noise, as is typical for ecological settings.

The observation by May (1974) that simple models of population growth can generate irregular oscillations in population abundance has made chaos of considerable interest among ecologists (Costantino et al. 1997; Perry et al. 2000; Cushing et al. 2003; Turchin 2003; Becks et al. 2005). Chaotically fluctuating populations may reach low densities, where their probability of extinction must be high (Thomas et al. 1980; Berryman and Millstein 1989). In this article we present a method that allows the regu-lation of chaotic population dynamics and that can prevent the population from going extinct.

* Corresponding author. Present address: Centre for Mathematical Biology, Mathematical and Statistical Sciences, CAB 501, University of Alberta, Edmonton, Alberta T6G 2G1, Canada; e-mail: fhilker@math.ualberta.ca.

† E-mail: frank.westerhoff@uos.de.

We emphasize that our aim is to regulate the chaotic system rather than to control it; that is, we avoid only certain unwanted fluctuations, namely, those in which the population would decline below its minimum viable size. The particularity of our approach is that the dynamics still remains chaotic, but it is less harmful. Unlike methods to suppress chaos (chaos control; e.g., Ott et al. 1990), the maintenance of chaos may be desirable for various reasons. For example, chaos has been shown to reduce the probability of global extinction in metapopulations (Allen et al. 1993; Ruxton 1994), to be an optimal regime for biological systems from an evolutionary perspective (Ferrière and Gatto 1993), and to favor biodiversity in species communities (Huisman and Weissing 1999).

The idea of our method is quite simple. In chaotic time series, one can often observe repeating crash patterns. Therefore, we first collect enough data to identify certain trajectories of population sizes that would lead to a crash. Then, we implement an intervention at these critical points to avoid bigger crashes in the near future (cf. so-called chaos anticontrol algorithms that can be traced back to Yang et al. 1995). The advantages of the method are obvious. The only requirement is time series information, which means that no knowledge about the underlying law of motion is needed. Furthermore, only occasional interventions of moderate size are sufficient, and we will show that they work well even for noisy and relatively short data sets, as is typical for ecological time series.

Belovsky et al. (1999) have demonstrated that nonlinear dynamics increases the likelihood of extinction in experimental populations of the brine shrimp (*Artemia franciscana*). Investigating time series of 758 species (including lepidopteran species, mammals, birds, and fish), Fagan et al. (2001) found that the majority of populations without immigrations or refuge mechanisms exhibit a tendency to fluctuate severely enough to possibly become extinct. There is a variety of mechanisms that especially threaten small populations (e.g., Soulé 1987; Lande 1988; Pimm et al. 1988): demographic stochasticity, environmental stochasticity, natural catastrophes, genetic inbreeding, edge effects (Lande 1987), antirescue effect (Harding and Mc-Namara 2002; Hovestadt and Poethke 2006), and the Allee effect (Allee 1931). This led to the concept of a minimum viable population, a deterministic threshold size below which a population is regarded to be particularly vulnerable to extinction (Shaffer 1981; Gilpin and Soulé 1986).

Because many species are increasingly threatened as a result of, for example, habitat fragmentation/degradation, biological invasions, and emerging or preexisting diseases, it is important to understand the dynamics of small populations as well as to explore effective management measures to preserve them (Lande 1988). Conservation biologists, for instance, have developed the framework of population viability analysis (see Boyce 1992) that estimates the future population size, assesses the extinction risk of an endangered species, and allows the exploration of appropriate management scenarios.

Chaotic population dynamics exhibits not only very small densities but also recurrent peaks in population numbers. These outbreaks continue to be a problem in both aquatic (e.g., algal blooms; Scheffer 1991; Hallegraeff 1993) and terrestrial ecosystems (e.g., insect outbreaks; Berryman 1987; Dwyer et al. 2004). Similarly, forest fires in the Mediterranean are predicted to be chaotic (Casagrandi and Rinaldi 1999), and epidemics of childhood diseases such as measles (e.g., Schaffer and Kot 1985; Olsen and Schaffer 1990; Drepper et al. 1994; Ellner et al. 1998) have been suggested to be "possibly chaotic" (Grenfell 2000, p. 49). Our method is general enough to also prevent those peaks of pest species or, more generally, any unwanted event, that is, an undesirable region in phase space. We demonstrate this by example with a model of the flour beetle *Tribolium castaneum* (Desharnais et al. 2001).

The remainder of this article is organized as follows. First, we introduce our method by applying it to time series data from the deterministic and stochastic Ricker map, test its robustness against the length of the available data sets, and investigate its effectiveness. Here, the aim is to prevent species extinction. Then we confront our method with an ecologically more realistic situation, namely, with a stage-structured population of the flour beetle. The aim is to prevent recurrent outbreaks. Finally, we discuss the applicability of this method as well as its limitations.

## The Method: Preventing Population Extinction

A typical time series of a chaotically fluctuating population is shown in figure 1a. It has been obtained by iterating the Ricker (1954) map as follows:

$$x_{t+1} = x_t \exp\left[r\left(1 - \frac{x_t}{K}\right)\right] =: f(x_t), \quad (1)$$

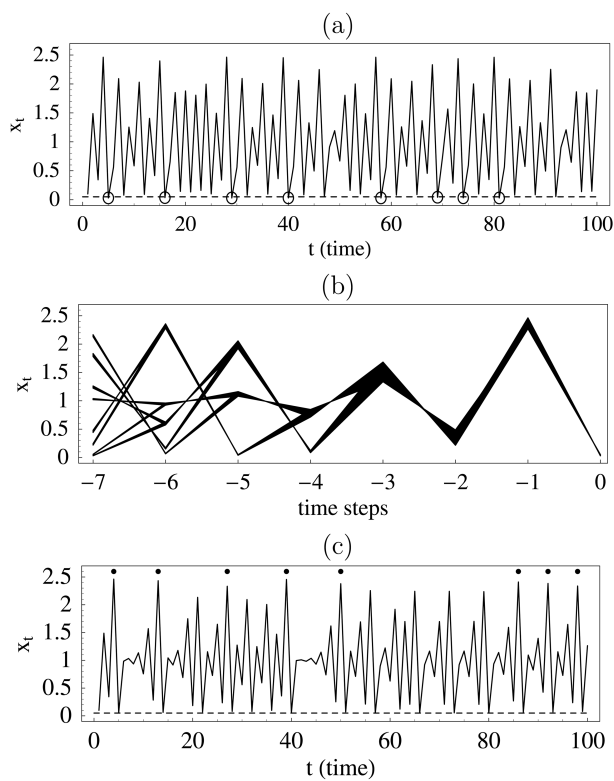

**Figure 1:** *a*, Chaotic time series generated by the Ricker map (eq. [1]) with $r = 3$, $K = 1$, $x_0 = 0.1$. The dashed line marks the threshold value $x^* = 0.05$. The circles indicate population crashes. *b*, Alert zones obtained from 10,000 observations (containing 867 crash paths). A crash is defined as a drop below the threshold value. *c*, Simulation of the dynamics with intervention as in equation (2) with $I = 0.25$. Each intervention is marked by a point in the top. The alert zone is $2.27232 \leq x_t \leq 2.46302$. Other parameters as in *a*.

where $x_t$ denotes population size (or density) at time $t$ and parameters $r > 0$ and $K > 0$ describe the intrinsic growth rate and carrying capacity, respectively. Assuming that there is a minimum viable population size, we set a threshold $x^*$ (fig. 1a, *dashed line*) below which we regard the population to be especially endangered. For an arbitrary value of $x^*$, we can see that the population in figure 1a falls below this critical value eight times (*circles*). In the long run (10,000 generations), the proportion of these crashes is 8.7%.

Because these crashes can be harmful, we define the interval of population states $0 \leq x_t \leq x^*$ as the undesirable (crash) region $U$ (termed "loss region" in Yang et al. 1995). Our method starts with finding the paths of trajectories that end in this crash region (risk trajectories). This allows us to implement an intervention mechanism that perturbs risk trajectories slightly so that they follow a different path.

*Step 1: Identifying Alert Zones*

Detecting risk trajectories in our method is straightforward: whenever the population $x_t$ falls into the crash region $U$, we simply scan the time series to collect the previous $m$ states $x_{t-1}, x_{t-2}, \ldots, x_{t-m}$. In other words, we trace the crash path by considering the successive preiterates of $x_t$. Repeating this for every crash that can be observed in the time series, we obtain a collection of crash paths as illustrated in figure 1b. For each time step $t - m$, there is a set of population states $Z_m$ that can be interpreted as alert zones: when the population is within these dangerous regions, the population is likely to crash in $m$ time steps. We can observe that the alert zones initially consist of single intervals but are then composed of an increasing number of smaller state intervals. Generally, for increasing $m$, the width of the alert zone tends to shrink in the unstable direction (as a consequence of the exponential divergence due to chaos; cf. Yang et al. 1995).

*Step 2: Implementing Interventions*

The identification of alert zones suggests that we proceed by fixing a value of $m$ and intervening each time the trajectory falls into an alert zone $Z_m$. By way of example, we present an intervention that takes place one time step before the identified crash (i.e., $m = 1$). This may be easy to implement in practice because we note that a crash is preceded by an immediate peak in population size (cf. fig. 1b). In ecological terms, this reflects scramble competition between individuals. We choose an intervention that directly affects the population state (e.g., by removing/harvesting/culling a constant number of individuals $I > 0$). The corresponding mathematical description is

$$x_{t+1} = \begin{cases} f(x_t - I) & \text{if } x_t \in Z_1 \\ f(x_t) & \text{otherwise} \end{cases}. \qquad (2)$$

In order to assess the effect of this intervention, figure 1c shows a simulation of equation (2) in which each intervention is marked by a point at the top. Population crashes are fully prevented; in a long-run simulation of 10,000 time steps, they do not occur a single time. This can be explained by taking a closer look at the dynamical map that results when incorporating the interventions. Figure 2 shows the realized map (eq. [2]; *solid line*) superimposed on the uncontrolled Ricker map (*dashed line*). One can readily see that the map in the alert zone is replaced by a part of the map that has been shifted from the left-hand side. As a result, the trajectory does not reach the crash region. The alert zone in figure 2 has been determined analytically, and the decisive lower bound of the exact alert zone (2.27215…) is very close to the one estimated from 10,000 observations (2.27232).

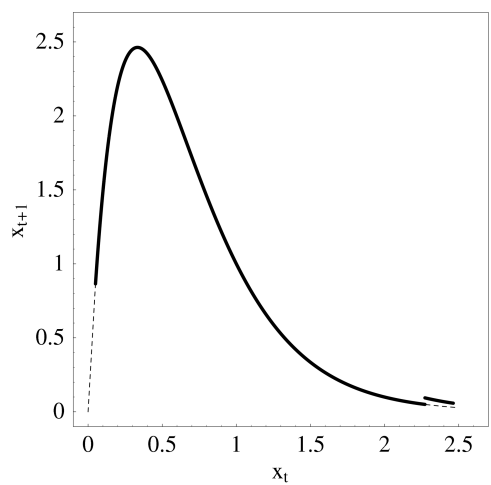

**Figure 2:** The intervention in the alert zone $x_t \in Z_1 \approx [2.27, 2.46]$ prevents the dynamics from reaching the crash region $x_{t+1} \in U = [0, x^* = 0.05]$. This map was obtained from 10,000 iterations of equation (2). The dashed line corresponds to the unperturbed Ricker map. Parameters as in figure 1.

When choosing intervention size $I$, we need to be careful for two reasons. First, $I$ needs to be sufficiently large to kick the trajectory out of the alert zone (and also not into the alert zone of another time step). Second, we will later see in a bifurcation diagram that too large a choice of $I$ may result in the suppression of chaotic dynamics, undesirable for the reasons stated in the introduction to this article. As a rule of thumb, we therefore choose $I$ slightly larger than the length of the alert zone. Because of the exponential decay of alert zone interval lengths, the required intervention sizes become smaller the earlier we intervene. Were it possible to implement long-sighted monitoring, alert zones would become rather thin, and one could exploit the sensitivity of chaotic dynamics on initial conditions, whereby only tiny interventions would be required. However, from the applied point of view, the unfolding of alert zones in several intervals complicates a straightforward management rule. Moreover, the method is probably more practical with later interventions, especially because noise might additionally perturb the intervened system in the meantime.

In principle, any intervention can be applied as long as it forces the trajectory to leave the crash path. Alternatively, knowledge of the crash paths can also be used to induce crashes or other phenomena (Hilker and Westerhoff 2007). For example, one could apply density-dependent interventions, which are widely used in sustainable harvesting in fisheries and forestry (e.g., Clark 1990). In "Robust-ness," we will point out that other interventions do not change the qualitative results and are also quantitatively very similar.

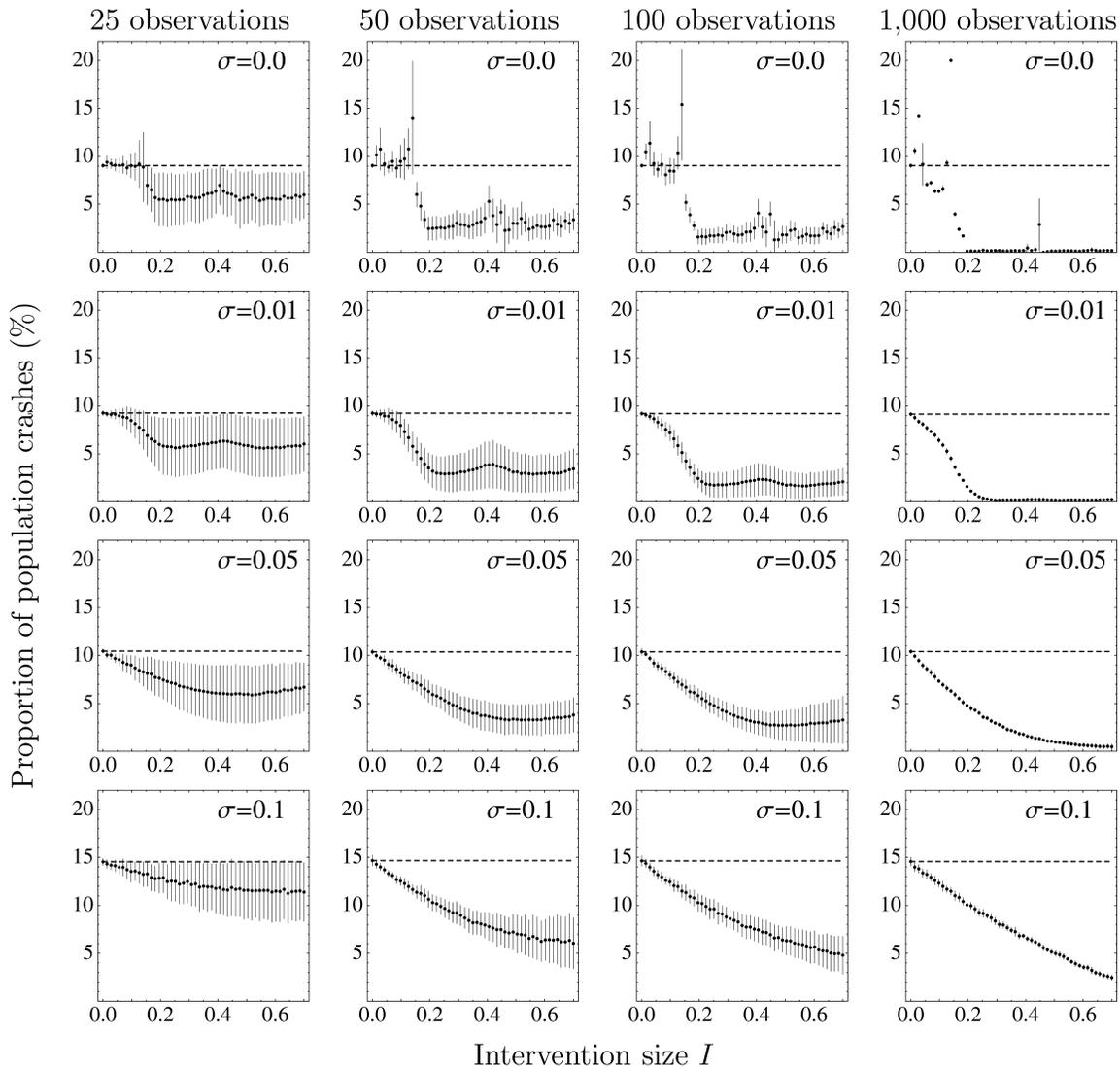

**Figure 3:** Effectiveness of the direct (culling) intervention method (eq. [2]) to prevent crashes in the general population model of the deterministic (*top row*) and stochastic Ricker map (eq. [3]) with different noise levels σ. The number of observations used to estimate the alert zone is indicated. Points give the mean effectiveness of 20 replicates and the vertical bars the standard deviation. The dotted line is the mean of the baseline scenario without intervention, that is, $I = 0$. Parameters: $r = 3$, $K = 1$.

### Effectiveness

Let us now explore the effectiveness of our intervention method in relation to the length of the time series. Generally, the more observations and crashes are present in the available time series, the more accurate the identification of alert zones and the less crashes are expected to occur. Again, we synthetically generate time series of various lengths in order to estimate the effectiveness of interventions and perform long-term simulations with the alert zones identified. In the top row of figure 3, we plot the proportion of population crashes in the simulation based on population size declining below the threshold value against intervention size $I$. From left to right, we use 25, 50, 100, and 1,000 observations to calculate the alert zones. The results for each value of $I$ are presented for 20 replicates with 10,000 simulation iterations each. Because initial conditions may influence both the proportion of crashes and the estimation of alert zones, we choose different initial conditions for each simulation. Baseline scenarios are shown for comparison ($I = 0$, equivalent to no intervention).

When intervention size $I$ exceeds the length of the alert zone (here approximately 0.19), the proportion of population crashes decreases significantly. For 25 observations, crash occurrences decreased from 9.0% to only 5.4%, which corresponds to a reduction of 40%. With an increasing number of observations, intervention performance continues to improve. For 50 and 100 observations, crashes can be reduced by 69% and 82%, respectively. Interventions based on 1,000 observations have the potential to prevent crashes almost completely.

While the control action can be adverse for subcritical interventions ($I \leq 0.19$), it is always beneficial for supercritical interventions ($I > 0.19$). In fact, the bifurcation diagram in figure 4 demonstrates that the undesirable region of population crashes, $U = [0, x^* = 0.05]$, is missing for intervention sizes $I > 0.19$. It also explains the poor performance around $I = 0.14$, where there is a low-period cycle with one of the population states corresponding to a crash. We can similarly explain why the effectiveness for supercritical interventions varies with size $I$. For imperfectly estimated alert zones, a small line of asymptotic values appears below $x^*$ in the bifurcation diagram (not shown here). Given that there is a periodic window around $I = 0.45$, the effectiveness is not as good as it is for other intervention sizes.

### Effect of Noise

Last, let us examine the effect of noise. Because we are primarily interested in the dynamics of small populations prone to extinction, we consider a demographic stochastic version of the Ricker map that adds a random variable $E_t$ on the square root scale

$$x_{t+1} = (\sqrt{f(x_t)} + E_t)^2, \qquad (3)$$

with $f(x_t)$ as in equation (1), $E_t$ being a normal random variable with mean 0 and a constant standard deviation (noise level) $\sigma$, and $E_0, E_1, E_2, \ldots$, being uncorrelated. In the case in which $[f(x_t)]^{1/2} + E_t$ is negative, we replace it with 0 (see Dennis et al. 2001; Cushing et al. 2003; Domokos and Scheuring 2004).

The proportion of population crashes observed for different noise levels is shown in separate rows in figure 3. First of all, we observe that interventions appear to be always beneficial—contrary to the deterministic case, even for subcritical intervention sizes ($I \leq 0.19$). Once more, we see that the number of population crashes can already be effectively reduced based on 25 observations. For small noise ($\sigma = 0.01$), the general pattern looks very similar to the deterministic case: (i) with a greater number of observations, the method becomes increasingly effective, and (ii) a critical intervention size approximately corresponds to the width of the alert zone. However, the lengths of alert zones become wider in the stochastic case. As a consequence, the sharp transition observed around $I = 0.19$ in the deterministic scenario becomes more fuzzy. Moreover, there is a larger variability in the effectiveness, but interventions continue to be always beneficial for supercritical $I$. For intermediate noise levels ($\sigma = 0.05$), the clear distinction between super- and subcritical intervention sizes vanishes. Intervention effectiveness seems to scale exponentially with $I$ (for long observations). In the presence of strong noise ($\sigma = 0.1$), crashes appear to decrease more linearly with intervention size. However, because alert zones become wider and wider with increasing noise variability, even large intervention sizes might fail to kick the trajectory out of the alert zone. Consequently, the "linearly" distributed effectiveness observed for large noise is the initial part of the scaling that in fact decreases more exponentially with $I$ (not shown here). Interestingly, intervention effectiveness appears to become less dependent on the number of observations available for stronger noise.

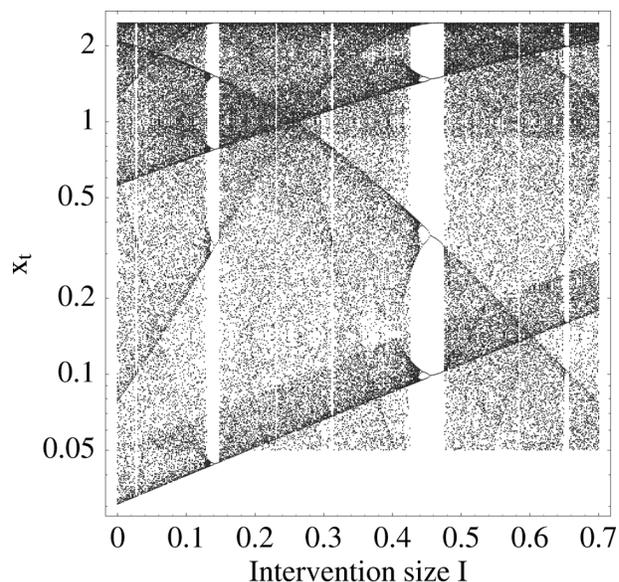

**Figure 4:** Bifurcation diagram of the deterministic Ricker map with direct culling intervention (eq. [2]) based on the alert zone $Z_1$ estimated from 10,000 observations. Note the logarithmic vertical scale for better visibility. Model parameters as in figure 1.

## Application to a Stage-Structured Insect Population: Preventing Outbreaks

In this section, we apply the previously described method to a well-parameterized and well-investigated three-dimensional population model of the flour beetle *Tribolium castaneum* (e.g., Costantino et al. 1997, 2005; Cushing et al. 2003). The model is stage structured and distinguishes among feeding larvae ($L_t$), nonfeeding larvae, pupae and callow adults ($P_t$), and sexually mature adults ($A_t$). Model equations with demographic stochasticity, henceforth called the LPA model, are as follows (Desharnais et al. 2001):

$$L_t = [\sqrt{bA_{t-1}\exp(-c_{EL}L_{t-1} - c_{EA}A_{t-1})} + E_{1t}]^2, \quad (4)$$

$$P_t = [\sqrt{L_{t-1}(1-\mu_L)} + E_{2t}]^2, \quad (5)$$

$$A_t = [\sqrt{P_{t-1}\exp(-c_{PA}A_{t-1}) + A_{t-1}(1-\mu_A)} + E_{3t}]^2, \quad (6)$$

where $b$ describes the number of larval recruits per adult and $\mu_L$ and $\mu_A$ describe the mortality of larvae and adults, respectively. The parameter $c_{EL}$ describes the cannibalism of eggs by larvae, $c_{EA}$ the cannibalism of eggs by adults, and $c_{PA}$ the cannibalism of pupae by adults. The vector $E_t = (E_{1t}, E_{2t}, E_{3t})'$ is made up of random noise terms with joint normal probability distribution with a mean vector of zeros and a variance-covariance matrix $\Sigma$. The deterministic version of system (4)–(6) can be obtained by setting $\Sigma = 0$ or, equivalently, $E_{1t} = E_{2t} = E_{3t} = 0$. For a certain range of parameter values, the deterministic model shows chaotic oscillations that have also been obtained experimentally (Costantino et al. 1997). Figure 5a shows the chaotic attractor of the deterministic LPA model, including distinct recurrent peaks of all state variables.

*Tribolium castaneum* is a significant agricultural pest. Therefore, we aim to prevent it from attaining those ranges of the chaotic attractor that correspond to mass outbreaks. In contrast to those described in the previous section, the undesirable regions are now large, instead of small, population sizes. However, we similarly confine desirable areas of the attractor by setting appropriate thresholds for the observable state variables. By way of example, we arbitrarily choose $A^* = 100$ and define system states $(L_t, P_t, A_t)$ with $A_t > A^*$ as outbreaks. The next step in our method is to learn the paths that lead to an outbreak, that is, to identify alert zones. Alert zones are now embedded in a three-dimensional state space. They actually turn out to be small bands of moderate length, as shown in figure 5b for the deterministic case.

A point where intervention can be applied is the time step that immediately precedes the next outbreak because the alert zone is very compact. We model the intervention by adding a certain number $I$ of adult individuals. This technique to manipulate laboratory populations has been used in the experiments by Desharnais et al. (2001). In order to assess the impact and effectiveness of this intervention, we again use the LPA model (4)–(6) to simulate the dynamics and record the resulting fraction of outbreaks. Results are shown in figure 6 for time series of various lengths. Without any interventions, an average of 12% of measurements correspond to outbreaks of the flour beetle. This fraction declines significantly for few available observations. Surprisingly, our method works for as few as 25 data points and can on average halve the number of mass outbreaks for 100 observations. For 1,000 observations, outbreaks can be consistently reduced to one-third. In particular, control actions generally lead to an improvement. The longer the time series available for es-timating the alert zone, the less variable and thus more reliable the interventions.

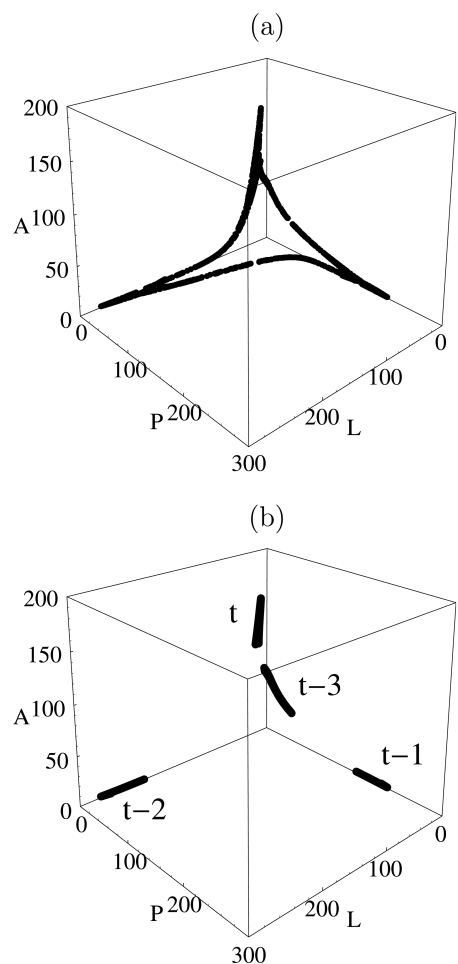

**Figure 5:** *a*, Chaotic attractor of the deterministic flour beetle model (eqq. [4]–[6]). *b*, Alert zones obtained from reconstructing outbreak paths in the state space leading to $A_t > A^*$, using 10,000 observations. Parameter values: $b = 10.45$, $c_{EL} = 0.01731$, $c_{EA} = 0.0131$, $\mu_L = 0.2$, $\mu_A = 0.96$, $c_{PA} = 0.35$, $E_t = 0$, $A^* = 100$.

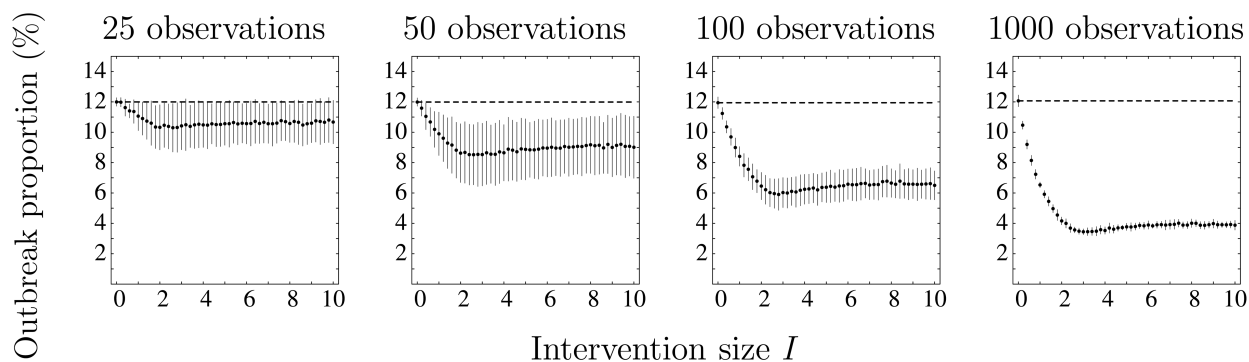

**Figure 6:** Effectiveness of the intervention method to prevent outbreaks of the stochastic flour beetle model (eqq. [4]–[6]). The outbreak proportion gives the fraction of states with $A_t > A^*$. The intervention size $I$ represents the number of adults added. The number of observations used to estimate the alert zone is indicated. Parameters as in figure 5 except $E_t \neq 0$ with $\Sigma_{11} = 2.332$, $\Sigma_{22} = 0.2364$, and all other entries of $\Sigma$ being equal to 0 (Desharnais et al. 2001). Each of the 20 replicates has been started with different initial condition.

When there are enough observations to estimate the alert zones accurately, it is possible to identify a critical value for intervention size around $I = 2$ above which intervention effectiveness reaches a plateau. Again, this critical intervention size is a good approximation to the interval length of the $A$ component in the estimated alert zones.

## Discussion

Chaotic fluctuations in population abundance may threaten ecosystem integrity by driving populations down (even leading to population extinctions) or up (corresponding to mass outbreaks with a multiplicity of possibly adverse effects). Controlling this situation seems to be intricate because of the erratic aspect of chaotic dynamics, but it is well known that even complex dynamics obeys some simple rules (e.g., Schaffer 1985; Schaffer and Kot 1986; Rinaldi et al. 2001a). We have presented an easy-to-implement method that can prevent both population crashes and outbreaks or, more generally, any undesirable region of a chaotic attractor. This method consists of two steps. First, a time series of the population is analyzed to identify paths that lead to crashes or outbreaks. This elucidates and defines alert zones: whenever population abundance reaches one of these areas, it is very likely to attain an undesirable size within the near future. The second step is to manipulate population abundance to force it out of the dangerous path that would otherwise lead to the harmful state. We found that there is a critical intervention size in the order of the alert zone length above which control actions substantially improve the desired performance.

### Robustness

Our method has been tested in intensive numerical simulations of a general one-dimensional population model (Ricker map). It turns out also to be very effective in a more realistic and complicated model of the flour beetle that has been well parameterized in the laboratory. The intervention effectiveness depends on both the number of available data and the intervention size. While the former is important for the approximation of alert zones, the latter determines how well the system can be forced to leave risk trajectories. We also investigated the impact of noise on intervention performance. On the one hand, it plays a constructive role in widening the alert zones. This minimizes the risk that potential crash paths are not identified. On the other hand, larger intervention sizes are required to push the system out of the alert zone.

The control policy presented in this article consists of removing/adding a constant number of individuals. However, we recognize that many sustainable harvesting strategies in the ecological literature make use of state-dependent control policies (e.g., Clark 1990). We therefore tested our method with the stochastic Ricker model and three alternative interventions (applied only if the population is situated in the alert zone):

$$x_{t+1} = f(x_t - E_1 x_t) \quad \text{if } x_t \in Z_1,$$
$$x_{t+1} = f(x_t) \exp(-E_2 x_t) \quad \text{if } x_t \in Z_2,$$
$$x_{t+1} = f(x_t) \exp(-E_3) \quad \text{if } x_t \in Z_2,$$

where $E_1$ and $E_2$ are measures for both the catching (harvesting) effort and catchability and $E_3$ is a similar control variable used in a model by Piccardi and Ghezzi (1997). The second and third control policies are implemented after population reproduction and are thus based on alert zones estimated two time steps ahead. It turns out that the results are qualitatively identical and also quantitatively very similar to figure 3. This is actually not surprising because the intervention is applied only when the system is in the alert zone and the important mechanism is then just to generate a perturbation of sufficient size. We find the usage of a constant intervention size instructive because it clearly demonstrates that the control has to be chosen in the size of the alert zone length.

The results have also been checked for lognormal multiplicative noise in the Ricker map, in order to account for environmental rather than demographic stochasticity (cf. Brännström and Sumpter 2006). Our findings are robust to this modification as well.

### Limitations

In this article, we have considered data from the one-dimensional Ricker map and the low-dimensional LPA system (which can, in fact, be described by a close to one-dimensional motion in a phase space reconstruction; cf. Takens 1981). Many, or even most, ecological systems in nature are, however, high dimensional, which greatly increases the demand for available observations. In those systems, our method will probably be difficult to apply as a result of a lack of data. Nevertheless, we have demonstrated that our method works for a complex system such as that described by the LPA model of the flour beetle *Tribolium*, which could be tested in laboratory experiments.

There are two other possible limitations of our method. First, the available data need to contain a certain amount of crashes or outbreaks so that one can identify a sufficient number of risk paths. The less frequent those events are, the longer the time series needs to be. To overcome this difficulty, one might modify risk threshold definitions or implement a stronger perturbation. Second, the implementation of this method will also face the problem of determining exact abundance or imperfect application of an intervention. In multispecies systems, for example, not all species might be accessible for observation or intervention.

### Efficiency

Ideally, the alert zone estimated from time series would be as large as necessary (to cover all potential risk paths) and as small as possible (to minimize the number of perturbations, e.g., for economical or conservationist reasons). The aim of our method is to prevent extinction or undesirable events such as outbreaks. We do not take into account the costs of intervention here because we assume that the undesirable scenarios are invaluable, catastrophic, or irreversible (species extinction, biodiversity loss, ecosystem disturbance, harvest disappearance, socioeconomic consequences). An intervention size of $I = 0.2$ can be highly efficient in preventing entire population crashes; this corresponds to less than 10% of the lower bound of the alert zone. In the flour beetle system, adding two adults prevents the system from boosting to more than 100 adults. Possible extensions of the presented method could consider a benefit-to-cost ratio and aspects from optimal control theory.

### Chaos Maintenance

The presented approach not only is highly effective in avoiding undesired events but also confines rather than controls the chaotic dynamics. However, the maintenance of chaos is not guaranteed for all intervention sizes; that is, the dynamics might get stabilized to a periodic oscillation or fixed point. Though this can be beneficial in some situations, this might be undesirable in others, as described in the introduction to this article. A simple refinement might solve this problem: varying intervention size over time so that the system maintains chaotic dynamics. Therefore, we outline two possible research directions. First, environmental and demographic noise could be used to smooth out sharp dynamical transitions and may induce chaos in parameter regions where there is no chaos otherwise (Crutchfield et al. 1982). Lai et al. (2003) have shown that there is a critical noise level above which there is an intermittent behavior with the trajectory visiting an unstable chaotic saddle that coexists in a periodic window. Second, there are a number of algorithms that can be used to sustain chaos (In et al. 1995; Nagai and Lai 1995; Parmananda and Eiswirth 1996) and that might be used for the current situation. However, if applied to practical ecological situations, they should be adapted to perform with small data sets only and to require neither any knowledge of the system dynamics nor the existence of certain dynamical regimes.

### Applicability and Related Approaches

The applicability of our intervention in the flour beetle system, that is, the addition of adult individuals, has already been experimentally demonstrated by Desharnais et al. (2001). These authors were interested in the general effect of perturbations and identified hot regions of the attractor by calculating the local derivatives of the model equations—information that is usually hardly available in practice. In contrast, the method presented here just needs sampled abundances. Experimental addition of adults, however, turned out to dramatically dampen the oscilla-tions. Further empirical evidence for successful interventions affecting population cycles has been provided recently in large-scale field experiments. The reduction of a certain amount of vertebrate predators (by live-trapping and removal of nest sites) prevented crashes of small rodents in Finland (Korpimäki and Norrdahl 1998). Similarly, the elimination of parasites in red grouse (by treatment with anthelminthics) prevented population crashes in northern England (Hudson et al. 1998).

Theoretical work that is related to this study is by Yang et al. (1995) and Dhamala and Lai (1999) as well as Shulenburger et al. (1999). Their attention, however, was focused on the case of transient chaos, that is, when the chaotic attractor has actually lost its stability. Piccardi and Ghezzi (1997) analyzed attractor confinement of the Ricker map from an optimal control perspective. Other approaches include peak-to-peak dynamics, applied to a forest-pest model (Rinaldi et al. 2001*b*), and symbolic time series analysis, applied to an epidemic problem (Piccardi 2004). However, the results presented here are, to our knowledge, the first systematic investigation in an ecologically applicable context. In particular, we take into account the amount of available data as well as varying intervention sizes. Our method is robust in the presence of noise and does not require any knowledge of the system equations. We ought to mention that all model equations appearing in this article are used only to generate time series and to simulate the effect of interventions. Thus, they serve only illustration purposes, and they are not necessary for the presented concept itself.

## Prospects

Notably, as few as 25 observations can yield sufficient information of the system to implement adequate interventions aimed to reduce flour beetle outbreaks. This amount of information is typically available in ecological time series. Full population viability analyses of endangered species, in contrast, continue to demand data sets that are only rarely available (e.g., Coulson et al. 2001). We therefore find the results of the presented approach encouraging to stimulate further refinements and to be tested in real situations. Our method might prove useful in resource management and contribute to improving regulation of population crashes, extinctions, and outbreaks, thus aiding areas as seemingly disparate as conservation biology, control of pest species, biological invasions, and risk analysis.

## Acknowledgments

We thank N. Corron, T. Hillen, M. Lewis, H. Malchow, N. Mantilla-Beniers, A. Potapov, and N. Stollenwerk for helpful comments and W. Schaffer and an anonymous reviewer for reviews. F.M.H. acknowledges support from the European Commission, grant MEXT-CT-2004-14338, and an Alberta Ingenuity and Honorary Killam postdoctoral fellowship.